\documentclass[
	aps,
	prl,
	reprint, % closely resembles final document format
	superscriptaddress,
	showpacs,
	showkeys,
	amsfonts,
	amsmath,
	amssymb,
]{revtex4-1}

\usepackage{hyperref}

\usepackage{multirow}
\usepackage{graphicx}

\usepackage{dcolumn}
\usepackage{bm}

\usepackage{gensymb}

\usepackage[capitalize]{cleveref}
\begin{document}

\title{Hard Two-Photon Contribution to Elastic Lepton-Proton
  Scattering\\Determined by the OLYMPUS Experiment}

\author{B.\,S.~Henderson}\affiliation{Massachusetts Institute of
  Technology, Cambridge, MA, USA}

\author{L.D.~Ice}\affiliation{Arizona State University, Tempe, AZ, USA}

\author{D.~Khaneft}\affiliation{Johannes Gutenberg-Universit\"at,
  Mainz, Germany}

\author{C.~O'Connor}\affiliation{Massachusetts Institute of
  Technology, Cambridge, MA, USA}

\author{R.~Russell}\affiliation{Massachusetts Institute of Technology,
  Cambridge, MA, USA}

\author{A.~Schmidt}\affiliation{Massachusetts Institute of Technology,
  Cambridge, MA, USA}

\author{J.\,C.~Bernauer}\email[Corresponding author:
]{bernauer@mit.edu}\affiliation{Massachusetts Institute of Technology,
  Cambridge, MA, USA}

\author{M.~Kohl} \email[Corresponding author:
]{kohlm@jlab.org}\altaffiliation{partially supported by Jefferson Lab}\affiliation{Hampton University, Hampton, VA, USA}

\author{N.~Akopov}\affiliation{Alikhanyan National Science Laboratory
  (Yerevan Physics Institute), Yerevan, Armenia}

\author{R.~Alarcon}\affiliation{Arizona State University, Tempe, AZ,
  USA}

\author{O.~Ates}
%\altaffiliation{Currently with St. Jude Children's
%  Research Hospital in Memphis, TN, USA}
\affiliation{Hampton
  University, Hampton, VA, USA}

\author{A.~Avetisyan}\affiliation{Alikhanyan National Science
  Laboratory (Yerevan Physics Institute), Yerevan, Armenia}

\author{R.~Beck}\affiliation{Rheinische
  Friedrich-Wilhelms-Universit\"at, Bonn, Germany}

\author{S.~Belostotski}\affiliation{Petersburg Nuclear Physics
  Institute, Gatchina, Russia}

\author{J.~Bessuille}\affiliation{Massachusetts Institute of
  Technology, Cambridge, MA, USA}

\author{F.~Brinker}\affiliation{Deutsches Elektronen-Synchrotron,
  Hamburg, Germany}

\author{J.\,R.~Calarco}\affiliation{University of New Hampshire, Durham,
  NH, USA}

\author{V.~Carassiti}\affiliation{Universit{\`a} degli Studi di Ferrara and Istituto
  Nazionale di Fisica Nucleare sezione di Ferrara, Ferrara, Italy}

\author{E.~Cisbani}\affiliation{Istituto Nazionale di Fisica Nucleare
  sezione di Roma and Istituto Superiore di Sanit\`a, Rome, Italy}

\author{G.~Ciullo}\affiliation{Universit{\`a} degli Studi di Ferrara and Istituto
  Nazionale di Fisica Nucleare sezione di Ferrara, Ferrara, Italy}\

\author{M.~Contalbrigo}\affiliation{Universit{\`a} degli Studi di Ferrara and
  Istituto Nazionale di Fisica Nucleare sezione di Ferrara, Ferrara,
  Italy}

%\author{N.~D'Ascenzo}\altaffiliation{Currently with Huazhong
%  University of Science and Technology, Wuhan, China and Institute of
%  Applied Mathematics, Russian Academy of Sciences,
%  Russia}\affiliation{Deutsches Elektronen-Synchrotron, Hamburg,
%  Germany}

\author{R.~De Leo}\affiliation{Istituto Nazionale di Fisica Nucleare
  sezione di Bari, Bari, Italy}

\author{J.~Diefenbach}
%\altaffiliation{Currently with Johannes
%  Gutenberg-Universit\"at, Mainz, Germany}
\affiliation{Hampton
  University, Hampton, VA, USA}

\author{T.\,W.~Donnelly}\affiliation{Massachusetts Institute of
  Technology, Cambridge, MA, USA}

\author{K.~Dow}\affiliation{Massachusetts Institute of Technology,
  Cambridge, MA, USA}

\author{G.~Elbakian}\affiliation{Alikhanyan National Science
  Laboratory (Yerevan Physics Institute), Yerevan, Armenia}

\author{P.\,D.~Eversheim}\affiliation{Rheinische
  Friedrich-Wilhelms-Universit\"at, Bonn, Germany}

\author{S.~Frullani}\affiliation{Istituto Nazionale di Fisica Nucleare
  sezione di Roma and Istituto Superiore di Sanit\`a, Rome, Italy}

\author{Ch.~Funke}\affiliation{Rheinische
  Friedrich-Wilhelms-Universit\"at, Bonn, Germany}

\author{G.~Gavrilov}\affiliation{Petersburg Nuclear Physics Institute,
  Gatchina, Russia}

\author{B.~Gl\"aser}\affiliation{Johannes Gutenberg-Universit\"at,
  Mainz, Germany}

\author{N.~G\"orrissen}\affiliation{Deutsches Elektronen-Synchrotron,
  Hamburg, Germany}

\author{D.\,K.~Hasell}\affiliation{Massachusetts Institute of Technology,
  Cambridge, MA, USA}

\author{J.~Hauschildt }\affiliation{Deutsches Elektronen-Synchrotron,
  Hamburg, Germany}

\author{Ph.~Hoffmeister}\affiliation{Rheinische
  Friedrich-Wilhelms-Universit\"at, Bonn, Germany}

\author{Y.~Holler}\affiliation{Deutsches Elektronen-Synchrotron,
  Hamburg, Germany}

\author{E.~Ihloff}\affiliation{Massachusetts Institute of Technology,
  Cambridge, MA, USA}

\author{A.~Izotov}\affiliation{Petersburg Nuclear Physics Institute,
  Gatchina, Russia}

\author{R.~Kaiser}\affiliation{University of Glasgow, Glasgow, United
  Kingdom}

\author{G.~Karyan} \altaffiliation{Also with Alikhanyan National
  Science Laboratory (Yerevan Physics Institute), Yerevan, Armenia}
\affiliation{Deutsches Elektronen-Synchrotron, Hamburg, Germany}

\author{J.~Kelsey}\affiliation{Massachusetts Institute of Technology,
  Cambridge, MA, USA}

\author{A.~Kiselev}
%\altaffiliation{Currently with Brookhaven National
%  Laboratory, Brookhaven, NY, USA}
\affiliation{Petersburg Nuclear
  Physics Institute, Gatchina, Russia}

\author{P.~Klassen}\affiliation{Rheinische
  Friedrich-Wilhelms-Universit\"at, Bonn, Germany}

\author{A.~Krivshich}\affiliation{Petersburg Nuclear Physics
  Institute, Gatchina, Russia}

\author{I.~Lehmann}\affiliation{University of Glasgow, Glasgow, United
  Kingdom}

\author{P.~Lenisa}\affiliation{Universit{\`a} degli Studi di Ferrara and Istituto
  Nazionale di Fisica Nucleare sezione di Ferrara, Ferrara, Italy}\

\author{D.~Lenz}\affiliation{Deutsches Elektronen-Synchrotron,
  Hamburg, Germany}

\author{S.~Lumsden}\affiliation{University of Glasgow, Glasgow, United
  Kingdom}

\author{Y.~Ma}
%\altaffiliation{Currently with RIKEN, Nishina Center,
%  Advanced Meson Science Laboratory, Japan}
\affiliation{Johannes
  Gutenberg-Universit\"at, Mainz, Germany}

\author{F.~Maas}\affiliation{Johannes Gutenberg-Universit\"at, Mainz,
  Germany}

\author{H.~Marukyan}\affiliation{Alikhanyan National Science
  Laboratory (Yerevan Physics Institute), Yerevan, Armenia}

\author{O.~Miklukho}\affiliation{Petersburg Nuclear Physics Institute,
  Gatchina, Russia}

\author{R.\,G.~Milner}\affiliation{Massachusetts Institute of Technology,
  Cambridge, MA, USA}

\author{A.~Movsisyan} \altaffiliation{Also with Universit{\`a} degli
  Studi di Ferrara and Istituto Nazionale di Fisica Nucleare sezione
  di Ferrara, Ferrara, Italy} \affiliation{Alikhanyan National Science
  Laboratory (Yerevan Physics Institute), Yerevan, Armenia}

\author{M.~Murray}\affiliation{University of Glasgow, Glasgow, United
  Kingdom}

\author{Y.~Naryshkin}\affiliation{Petersburg Nuclear Physics
  Institute, Gatchina, Russia}

\author{R.~Perez~Benito}\affiliation{Johannes Gutenberg-Universit\"at,
  Mainz, Germany}

\author{R.~Perrino}\affiliation{Istituto Nazionale di Fisica Nucleare
  sezione di Bari, Bari, Italy}

\author{R.\,P.~Redwine}\affiliation{Massachusetts Institute of
  Technology, Cambridge, MA, USA}

\author{D.~Rodr\'iguez~Pi\~neiro}\affiliation{Johannes Gutenberg-Universit\"at, Mainz, Germany}

\author{G.~Rosner}\affiliation{University of Glasgow, Glasgow, United
  Kingdom}

\author{U.~Schneekloth}\affiliation{Deutsches Elektronen-Synchrotron,
  Hamburg, Germany}

\author{B.~Seitz}\affiliation{University of Glasgow, Glasgow, United
  Kingdom}

\author{M.~Statera}\affiliation{Universit{\`a} degli Studi di Ferrara
  and Istituto Nazionale di Fisica Nucleare sezione di Ferrara,
  Ferrara, Italy}\

\author{A.~Thiel}\affiliation{Rheinische
  Friedrich-Wilhelms-Universit\"at, Bonn, Germany}

\author{H.~Vardanyan}\affiliation{Alikhanyan National Science
  Laboratory (Yerevan Physics Institute), Yerevan, Armenia}\

\author{D.~Veretennikov}\affiliation{Petersburg Nuclear Physics
  Institute, Gatchina, Russia}

\author{C.~Vidal}\affiliation{Massachusetts Institute of Technology,
  Cambridge, MA, USA}

\author{A.~Winnebeck}
%\altaffiliation{Currently with Varian Medical
%  Systems, Bonn, Germany}
\affiliation{Massachusetts Institute of
  Technology, Cambridge, MA, USA}

\author{V.~Yeganov }\affiliation{Alikhanyan National Science
  Laboratory (Yerevan Physics Institute), Yerevan, Armenia}

\collaboration{The OLYMPUS Collaboration}\noaffiliation{}

\date{\today}

\begin{abstract}
  The OLYMPUS collaboration reports on a precision measurement of the
  positron-proton to electron-proton elastic cross section ratio,
  $R_{2\gamma}$, a direct measure of the contribution of hard
  two-photon exchange to the elastic cross section. In the OLYMPUS
  measurement, 2.01~GeV electron and positron beams were directed
  through a hydrogen gas target internal to the DORIS storage ring at
  DESY. A toroidal magnetic spectrometer instrumented with drift
  chambers and time-of-flight scintillators detected elastically
  scattered leptons in coincidence with recoiling protons over a
  scattering angle range of $\approx 20\degree$ to $80\degree$. The
  relative luminosity between the two beam species was monitored using
  tracking telescopes of interleaved GEM and MWPC detectors at
  $12\degree$, as well as symmetric M{\o}ller/Bhabha calorimeters at
  $1.29\degree$.  A total integrated luminosity of 4.5~fb$^{-1}$ was
  collected. In the extraction of $R_{2\gamma}$, radiative effects
  were taken into account using a Monte Carlo generator to simulate
  the convolutions of internal bremsstrahlung with experiment-specific
  conditions such as detector acceptance and reconstruction
  efficiency. The resulting values of $R_{2\gamma}$, presented here
  for a wide range of virtual photon polarization
  $0.456<\epsilon<0.978$, are smaller than some hadronic two-photon
  exchange calculations predict, but are in reasonable agreement with
  a subtracted dispersion model and a phenomenological fit to the form
  factor data.
\end{abstract}

\pacs{25.30.Bf 25.30.Hm 13.60.Fz 13.40.Gp 29.30.-h}

\keywords{elastic electron scattering; elastic positron scattering;
  two-photon exchange; form factor ratio}

\maketitle

Measurements of the proton's elastic form factor ratio,
$\mu_p G^p_E / G^p_M$, using polarization techniques~\citep{
  Hu:2006fy, MacLachlan:2006vw, Gayou:2001qt, Punjabi:2005wq,
  Jones:2006kf, Puckett:2010ac, Paolone:2010qc,Puckett:2011xg} show a
dramatic discrepancy with the ratio obtained using the traditional
Rosenbluth technique in unpolarized cross section
measurements~\citep{Litt:1969my, Bartel:1973rf, Andivahis:1994rq,
  Walker:1993vj, Christy:2004rc, Qattan:2004ht}.  One hypothesis for
the cause of this discrepancy is a contribution to the cross section
from hard two-photon exchange (TPE), which is not included in standard
radiative corrections and would affect the two measurement techniques
differently~\citep{ Guichon:2003qm, Blunden:2003sp, Chen:2004tw,
  Afanasev:2005mp, Blunden:2005ew, Kondratyuk:2005kk}. Standard
radiative correction prescriptions account for two-photon exchange
only in the soft limit, in which one photon carries negligible
momentum~\citep{Mo:1968cg, Maximon:2000hm}. There is no
model-independent formalism for calculating hard TPE.  Some
model-dependent calculations suggest that TPE is responsible for the
form factor discrepancy~\citep{Chen:2004tw, Afanasev:2005mp,
  Blunden:2005ew, Kondratyuk:2005kk} while others contradict that
finding~\citep{Bystritskiy:2006ju, Kuraev:2007dn}.

Hard TPE can be quantified from a measurement of $R_{2\gamma}$, the
ratio of positron-proton to electron-proton elastic cross sections
that have been corrected for the standard set of radiative effects,
including soft TPE. The interference of one- and two-photon exchange
is odd in the sign of the lepton charge, so any deviation in
$R_{2\gamma}$ from unity can be attributed to hard TPE. The OLYMPUS
experiment, as well as two recent experiments at
VEPP-3~\citep{Rachek:2014fam} and CLAS~\citep{Adikaram:2014ykv}, have
measured $R_{2\gamma}$ to specifically determine if hard TPE is
sufficient to explain the observed discrepancy in the proton's form
factor ratio, or if some additional explanation is needed.

Both the magnitude of $R_{2\gamma}$ and its kinematic dependence are
relevant. If hard TPE is the cause of the discrepancy,
phenomenological models~\citep{Chen:2007ac, Guttmann:2010au,
  Bernauer:2013tpr, Schmidt:2016aa} predict $R_{2\gamma}$ should rise
with decreasing $\epsilon$ and increasing $Q^2$. Here, $\epsilon$ is
the virtual photon polarization parameter given by
$[1+2(1+\tau)\tan^2(\theta_e/2)]^{-1}$, where $\theta_e$ is the lepton
scattering angle and $\tau = Q^2/(4M_p^2)$, where $M_p$ is the proton
mass, and $Q^2=-q_{\mu}q^{\mu}$ is the negative four-momentum transfer
squared.

Only a brief overview of the OLYMPUS experiment is given here
(see~\citep{Milner:2013daa} for a detailed description).  The OLYMPUS
experiment took data in the last running of the DORIS
electron/positron storage ring at DESY, Hamburg, Germany.  The DORIS
magnet power supplies were modified to allow the beam species to be
changed daily.  The experiment collected a total integrated luminosity
of $4.5$~fb$^{-1}$.  The 2.01~GeV stored beams with up to 65~mA of
current passed through an internal, unpolarized hydrogen gas target
with an areal density of approximately
$3\times10^{15}$~atoms/cm$^2$~\citep{Bernauer:2014pva}.

The detector was based on the former MIT-Bates BLAST
detector~\citep{Hasell:2009zza}: a toroidal magnetic spectrometer with
the two horizontal sections instrumented with large acceptance
($20\degree<\theta<80\degree$, $-15\degree<\phi<15\degree$) drift
chambers (DC) for 3D particle tracking and walls of time-of-flight
scintillator bars (ToF) for triggering and particle identification.
To a good approximation, the detector system was left-right symmetric
and this was used as a cross-check in the analysis.  Most of the data
were collected with positive toroid polarity to avoid excessive noise
rates in the DC due to low-energy electrons being bent away from the
beam axis into the DCs.

Two new detector systems were designed and built to monitor the
luminosity. These were symmetric M{\o}ller/Bhabha calorimeters (SYMB)
at $1.29\degree$~\citep{Benito:2016cmp} and two telescopes of three
triple gas electron multiplier (GEM) detectors~\citep{Ates:2014aa}
interleaved with three multi-wire proportional chambers (MWPC)
mounted at $12\degree$.

The trigger system selected candidate events that resulted from a
lepton and proton detected in coincidence in opposite sectors.  The
CBELSA/TAPS data acquisition system~\citep{Thiel:2012yj} was used to
readout the data and stored it to disk.

An optical survey of all detector positions was made and the magnetic
field was mapped throughout the tracking
volume~\citep{Bernauer:2016cc}.

A complete Monte Carlo (MC) simulation of the experiment was developed
in order to account for the differences between electrons and
positrons with respect to radiative effects, changing beam position
and energy, the spectrometer acceptance, track reconstruction
efficiency, luminosity, and elastic event selection.  Rather than
correct each effect individually, the simulation allowed the complete
forward propagation of the correlations amongst all of these
effects. The ratio we report is given by
\begin{equation}
  R_{2\gamma}= \left[
    \frac{N_\text{exp}(e^+)}{N_\text{exp}(e^-)}\right]\bigg/
  \left[ \frac{N_\text{MC}(e^+)}{N_\text{MC}(e^-)}\right]=
  \frac{R^\text{exp}}{R^\text{MC}},
\end{equation}
where $N_i$ are the observed and simulated counts.

The first stage in the simulation was a radiative event generator
developed specifically for
OLYMPUS~\citep{Russell:2016aa,Schmidt:2016aa}. This generator produced
lepton-proton events weighted by several different radiative cross
section models. In this letter, the results from four prescriptions
are presented: following Mo-Tsai~\citep{Mo:1968cg} and
Maximon-Tjon~\citep{Maximon:2000hm}, both with radiative effects to
order $\alpha^3$ and to all orders through exponentiation. The Mo-Tsai
order $\alpha^3$ prescription is equivalent to the ESEPP
generator~\citep{Gramolin:2014pva} used by the VEPP-3 experiment.  The
difference in $R_{2\gamma}$ extracted using the four approaches is as
much as 1.5\% at low $\epsilon$, indicating that higher-order effects
in radiative corrections are significant and depend on the effective
cutoff energy.

Particle trajectories were simulated using a three-dimensional model
of the apparatus and then digitized to produce simulated data in
exactly the same format as the experimental data. This digitization
procedure accounted for the efficiency and resolution of individual
detector elements, determined using data-driven approaches. Both the
experimental and simulated data were analyzed with the same analysis
code.

Track reconstruction was performed using a pattern matching
procedure on detector signals to identify track candidates.  Then two
distinct tracking algorithms were employed to fit the track initial
conditions: momentum, scattering angles, and vertex position.

Four independent elastic event selection routines were
developed~\citep{Henderson:2016aa, Russell:2016aa, Schmidt:2016aa,
  Bernauer:analysis}, and the results presented are the average of the
four with the statistical uncertainty calculated as the average of the
statistical uncertainty of each analysis. Two additional routines are
in preparation~\citep{O'Connor:2016aa, Ice:2016aa}. Each routine uses
different approaches, but all leverage the fact that the kinematics of
elastic events is over-determined so that cuts on reconstructed
kinematic quantities---momenta, angles, time-of-flight, vertex
positions of the lepton and proton---could be used to reduce
background from the sample of elastic events. Time-of-flight was used
effectively to discriminate leptons from protons. Cuts on the proton
acceptance were used to avoid acceptance edge effects. All of the
routines utilized a background subtraction procedure, and all
confirmed that the background rates were similar for electron and
positron modes. Background typically varied from negligible at low
$Q^2$ to $\approx20$\% at high $Q^2$.  The routines binned elastic
events according to the reconstructed proton angle, as this
reconstruction was identical in electron and positron modes.  We
report results on a subset of the total recorded data selected for
optimal running conditions, corresponding to $3.1$~fb$^{-1}$ of
integrated luminosity.

The integrated luminosity for each beam species was independently
monitored using the $12\degree$ telescopes, the SYMB, and from the
beam current and target density recorded by the slow control
system. The most accurate determination came from an analysis of
multi-interaction events (MIE) in the
SYMB~\citep{Schmidt:2016aa,Schmidt:MIE}. In the MIE analysis, the
luminosity was extracted from the ratio of rates of two types of
events. The denominator was the rate of symmetric M{\o}ller/Bhabha
events, in which the two final state leptons entered the SYMB. The
numerator was the rate of events in which a beam lepton scattered
elastically from a target proton and entered the calorimeter in random
coincidence with a symmetric M{\o}ller/Bhabha interaction. By
extracting the luminosity from a ratio of rates, the MIE analysis
exploited cancelations of several systematic uncertainties (like
acceptance, detector efficiency, etc.), reducing the uncertainty in
the relative luminosity between beam species to 0.36\%.

The redundant pair of tracking telescopes at $12\degree$ measured
elastic $ep$ scattered leptons in coincidence with recoil protons in
the DC and ToF around $72\degree$, which permitted a measurement of
$R_{2\gamma}$ at $\epsilon=0.978$ with negligible statistical
uncertainty using the MIE luminosity~\citep{Henderson:2016aa}.

\Cref{tab:systematics}
\begin{table}
  \caption{\label{tab:systematics}Contributions to the systematic
    uncertainty in $R_{2 \gamma}$.}
\begin{tabular}{l|c}
 \hline\hline
Correlated contributions & Uncertainty in $R_{2 \gamma}$\\
       \hline
Beam energy & 0.04--0.13\%\\
MIE luminosity & 0.36\% \\
Beam and detector geometry & 0.25\% \\
\hline\hline
Uncorrelated contributions\\
\hline
Tracking efficiency & 0.20\% \\
 \begin{tabular}{@{}l@{}}
Elastic selection and \\ $\;\;$background subtraction
\end{tabular} & 0.25--1.17\%\\
\hline\hline
\end{tabular}
\end{table}
summarizes the dominant contributions to the systematic uncertainty in
$R_{2 \gamma}$. The uncertainty from geometry was estimated from the
differences between $R_{2\gamma}$ extracted from left-lepton versus
right-lepton events. The uncertainty from tracking efficiency was
estimated from the performance of the two different tracking
algorithms. The uncertainty from elastic selection was estimated from
the variance in $R_{2\gamma}$ produced by the different selection
routines.

\begin{figure}[!ht]
\centering\includegraphics[width=\columnwidth]{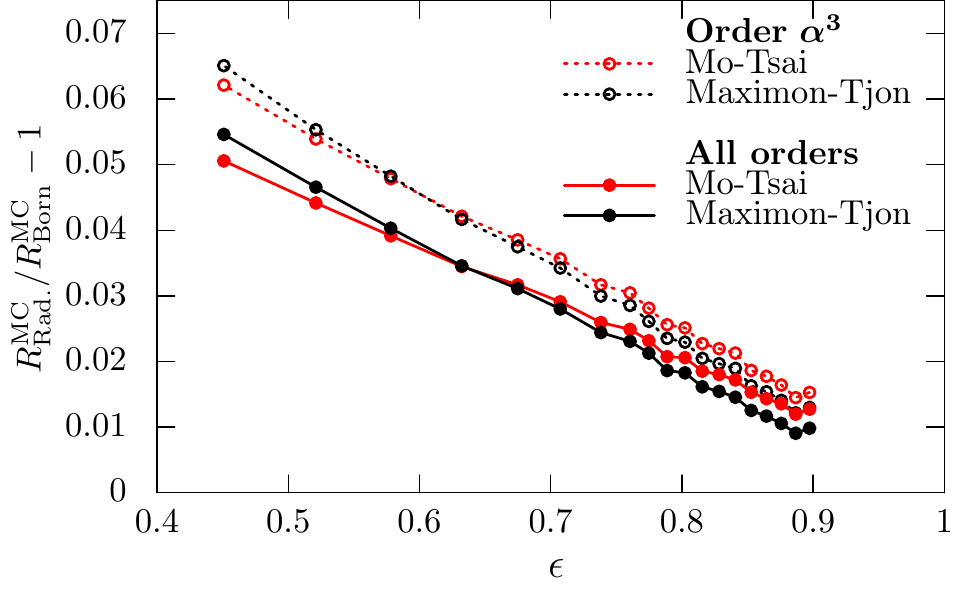}
\caption{Approximate effects of radiative corrections versus
  $\epsilon$ are on the order of several percent.}
\label{radcorr}
\end{figure}

We want to emphasize that radiative corrections have a large effect on
the OLYMPUS determination of $R_{2\gamma}$. The corrections to
$R_{2\gamma}$ are driven by the lepton charge-odd corrections: soft
TPE and the interference of bremsstrahlung off the lepton and
proton. In the OLYMPUS analysis, radiative effects cannot be unfolded
from the effects of detector efficiency, acceptance, etc., but the
magnitude of radiative effects on $R_{2\gamma}$ can be estimated by
comparing the full simulation with one where the events are
re-weighted by the first Born approximation. \Cref{radcorr}
shows the size of the correction for the four different prescriptions.
We find that the corrections are approximately 5--6\% at the lowest
$\epsilon$ values, and, furthermore, that higher-order effects can
alter the correction by as much as 1\%. The effective energy cut-off
in the analysis is only a few percent of the outgoing lepton
energy. In that range, the exponentiation should yield a more accurate
result.  The large dependency on the prescription used underscores
that theoretical improvements to the treatment of higher order
bremsstrahlung are crucial for future high-precision experiments.

\begin{figure}[!ht]
\centering\includegraphics[width=\columnwidth]{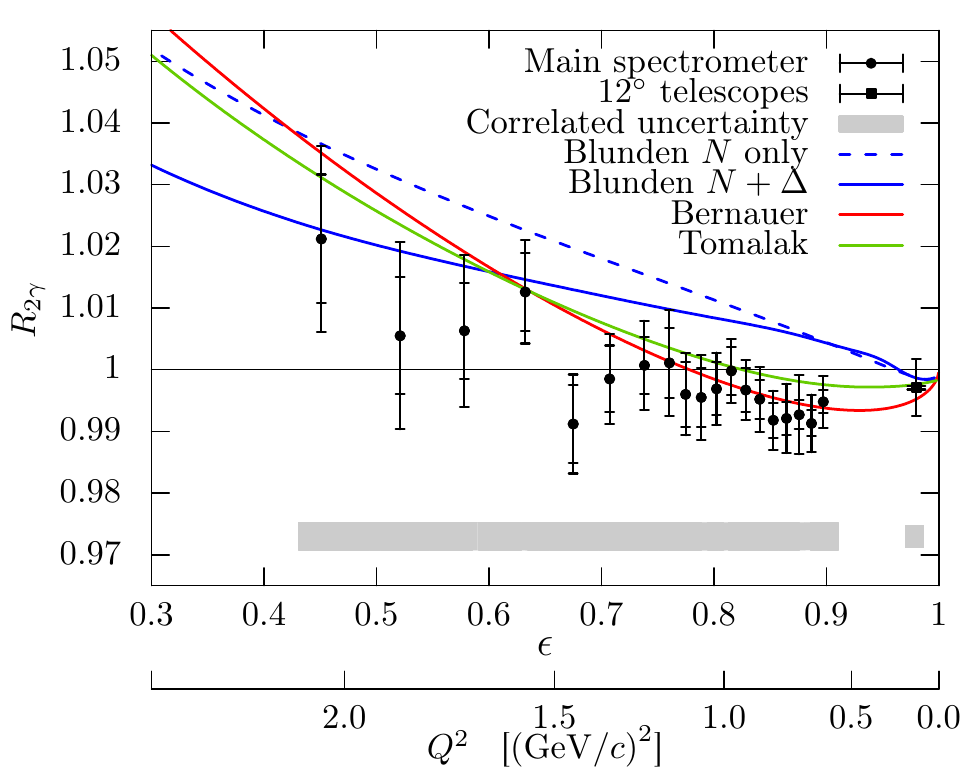}
\caption{ OLYMPUS result for $R_{2\gamma}$ using the
  Mo-Tsai~\citep{Mo:1968cg} prescription for radiative corrections to
  all orders. Uncertainties shown are statistical (inner bars),
  uncorrelated systematic (added in quadrature, outer bars), and
  correlated systematic (gray band). Note the $12\degree$ data point
  at $\epsilon=0.978$ is completely dominated by systematic
  uncertainties.}
\label{theratior}
\end{figure}

The OLYMPUS determination of $R_{2 \gamma}$ as a function of
$\epsilon$ and $Q^2$ is provided in \cref{tab:results}
\begin{table}
\begin{tabular}{c|c|c|c|c|c|c|c|c}
\hline \hline
$\langle\epsilon\rangle $ & $\langle Q^2\rangle $ & $R_{2\gamma}$ & $R_{2\gamma}$ &$R_{2\gamma}$ &$R_{2\gamma}$ &
$\delta_\text{stat.}$ & $\delta_\text{syst.}^\text{uncorr}$ & $\delta_\text{syst.}^\text{corr}$ \\
& $\frac{\text{GeV}^2}{c^2}$ & (a) & (b) & (c) & (d) & \multicolumn{3}{c}{$\times10^{-4}$}\\
\hline
0.978 & 0.165 & 0.9971 & 0.9967 & 0.9979 & 0.9978 & 3 & 46 & 36 \\
0.898 & 0.624 & 0.9920 & 0.9948 & 0.9944 & 0.9958 & 19 & 37 & 45 \\
0.887 & 0.674 & 0.9888 & 0.9913 & 0.9912 & 0.9923 & 21 & 42 & 45 \\
0.876 & 0.724 & 0.9897 & 0.9927 & 0.9921 & 0.9935 & 23 & 60 & 45 \\
0.865 & 0.774 & 0.9883 & 0.9921 & 0.9907 & 0.9929 & 26 & 50 & 45 \\
0.853 & 0.824 & 0.9879 & 0.9918 & 0.9903 & 0.9926 & 29 & 39 & 45 \\
0.841 & 0.874 & 0.9907 & 0.9952 & 0.9931 & 0.9958 & 32 & 42 & 45 \\
0.829 & 0.924 & 0.9919 & 0.9967 & 0.9943 & 0.9972 & 36 & 33 & 45 \\
0.816 & 0.974 & 0.9950 & 0.9998 & 0.9973 & 1.0002 & 39 & 33 & 45 \\
0.803 & 1.024 & 0.9913 & 0.9969 & 0.9936 & 0.9971 & 43 & 40 & 45 \\
0.789 & 1.074 & 0.9905 & 0.9955 & 0.9927 & 0.9956 & 47 & 50 & 45 \\
0.775 & 1.124 & 0.9904 & 0.9960 & 0.9926 & 0.9960 & 52 & 41 & 45 \\
0.761 & 1.174 & 0.9950 & 1.0011 & 0.9971 & 1.0009 & 57 & 63 & 45 \\
0.739 & 1.246 & 0.9945 & 1.0007 & 0.9964 & 1.0002 & 46 & 56 & 45 \\
0.708 & 1.347 & 0.9915 & 0.9985 & 0.9930 & 0.9977 & 54 & 49 & 46 \\
0.676 & 1.447 & 0.9842 & 0.9912 & 0.9854 & 0.9899 & 63 & 50 & 46 \\
0.635 & 1.568 & 1.0043 & 1.0126 & 1.0049 & 1.0105 & 63 & 55 & 46 \\
0.581 & 1.718 & 0.9968 & 1.0063 & 0.9966 & 1.0032 & 77 & 96 & 46 \\
0.524 & 1.868 & 0.9953 & 1.0055 & 0.9941 & 1.0013 & 95 & 118 & 46 \\
0.456 & 2.038 & 1.0089 & 1.0212 & 1.0064 & 1.0154 & 104 & 108 & 46 \\
\hline\hline
\end{tabular}
\caption{\label{tab:results}OLYMPUS results for $R_{2\gamma}$
  using the prescriptions: Mo-Tsai to order $\alpha^3$ (a) and to all
  orders (b); and using Maximon-Tjon to order $\alpha^3$ (c) and to all
  orders (d).}
\end{table}
for the four different radiative correction prescriptions. The results
using Mo-Tsai to all orders are shown in \cref{theratior},
along with theoretical calculations by Blunden~\citep{Blunden:2016aa}
and by Tomalak~\citep{Tomalak:2014sva} and the phenomenological
prediction from Bernauer's~\citep{Bernauer:2013tpr} fit to unpolarized
and polarized proton form factors measurements that includes a
parameterization for the TPE $\epsilon$ and $Q^2$ dependence.  OLYMPUS
finds that the contribution from hard TPE is small at this beam energy
though there is a noticeable trend from below unity at higher values
of $\epsilon$ increasing to around 2\% at $\epsilon=0.46$.  The
results are in general below the theoretical prediction of Blunden.
The subtracted dispersion calculation of Tomalak shown has used
Bernauer's form factor data and a subtraction point at $\epsilon=0.5$.
Both Tomalak's calculation and Bernauer's phenomenological prediction
are in reasonable agreement with the OLYMPUS results.

\begin{figure}[!ht]
\centering\includegraphics[width=\columnwidth]{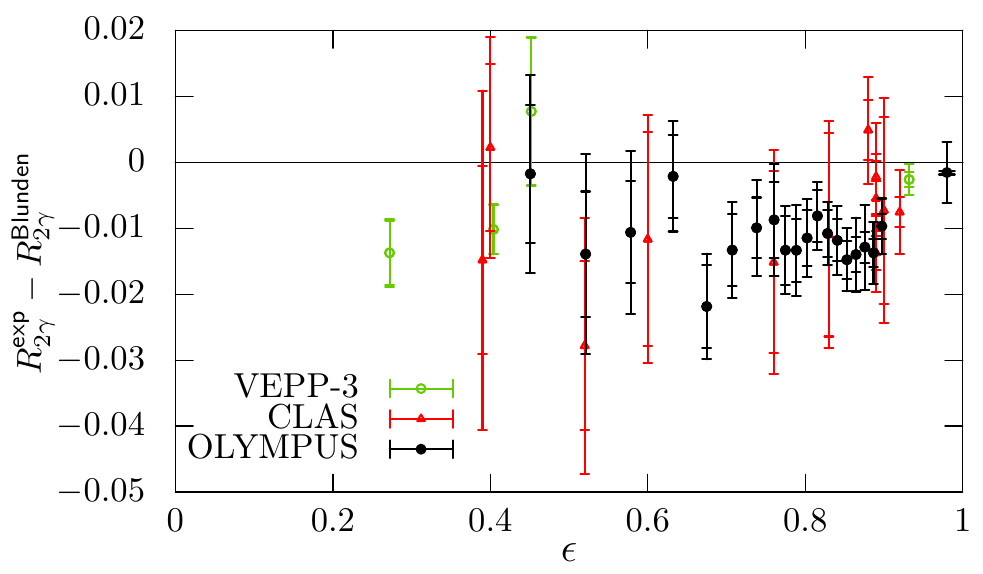}
\caption{ Comparison of the recent results to the calculation by
  Blunden.  The data are in good agreement, but generally fall below
  the prediction. Please note that data at similar $\epsilon$ values
  have been measured at different $Q^2$. Also note that the VEPP-3
  data have been normalized to the calculation at high $\epsilon$.}
\label{diffblunden}
\end{figure}

A comparison of the results from recent $R_{2\gamma}$ experiments to
Blunden's newest calculation ($N+\Delta$) is shown in
\cref{diffblunden}.  We plot the difference between the data and
theory calculated at the $\epsilon$ and $Q^2$ for each data point to
approximately take into account that the data were taken at different
$\epsilon$ and $Q^2$ values.  This shows the data are largely
consistent with each other, but mostly below the calculation by
Blunden.  A similar plot could be made versus $Q^2$.  Comparison with
the phenomenological prediction of Bernauer (not shown) shows good
agreement.

We do not agree with the conclusions of the earlier
papers~\citep{Adikaram:2014ykv, Rachek:2014fam}.  The data shown in
\cref{diffblunden} clearly favours a smaller $R_{2\gamma}$.  While the
agreement with the phemonological prediction of Bernauer suggests that
TPE is causing most of the discrepancy in the form factor ratio in the
measured range. The theoretical calculation of Blunden, which shows
roughly enough strength to explain the discrepancy at larger $Q^2$,
does not match the data in this regime.  To clarify the situation, the
size of TPE at large $Q^2$ has to be determined in future
measurements.

\begin{acknowledgments}
  We gratefully acknowledge P.G. Blunden for the theoretical
  calculations used in this letter.  We thank the DORIS machine group
  and the various DESY groups that made this experiment possible.  We
  gratefully acknowledge the numerous funding agencies: the Ministry
  of Education and Science of Armenia, the Deutsche
  Forschungsgemeinschaft, the European Community-Research
  Infrastructure Activity, the United Kingdom Science and Technology
  Facilities Council and the Scottish Universities Physics Alliance,
  the United States Department of Energy and the National Science
  Foundation, and the Ministry of Education and Science of the Russian
  Federation. R. Milner also acknowledges the generous support of the
  Alexander von Humboldt Foundation, Germany.
\end{acknowledgments}

\appendix

\bibliography{tpeprl}

\begin{thebibliography}{46}%
\makeatletter
\providecommand \@ifxundefined [1]{%
 \@ifx{#1\undefined}
}%
\providecommand \@ifnum [1]{%
 \ifnum #1\expandafter \@firstoftwo
 \else \expandafter \@secondoftwo
 \fi
}%
\providecommand \@ifx [1]{%
 \ifx #1\expandafter \@firstoftwo
 \else \expandafter \@secondoftwo
 \fi
}%
\providecommand \natexlab [1]{#1}%
\providecommand \enquote  [1]{``#1''}%
\providecommand \bibnamefont  [1]{#1}%
\providecommand \bibfnamefont [1]{#1}%
\providecommand \citenamefont [1]{#1}%
\providecommand \href@noop [0]{\@secondoftwo}%
\providecommand \href [0]{\begingroup \@sanitize@url \@href}%
\providecommand \@href[1]{\@@startlink{#1}\@@href}%
\providecommand \@@href[1]{\endgroup#1\@@endlink}%
\providecommand \@sanitize@url [0]{\catcode `\\12\catcode `\$12\catcode
  `\&12\catcode `\#12\catcode `\^12\catcode `\_12\catcode `\%12\relax}%
\providecommand \@@startlink[1]{}%
\providecommand \@@endlink[0]{}%
\providecommand \url  [0]{\begingroup\@sanitize@url \@url }%
\providecommand \@url [1]{\endgroup\@href {#1}{\urlprefix }}%
\providecommand \urlprefix  [0]{URL }%
\providecommand \Eprint [0]{\href }%
\providecommand \doibase [0]{http://dx.doi.org/}%
\providecommand \selectlanguage [0]{\@gobble}%
\providecommand \bibinfo  [0]{\@secondoftwo}%
\providecommand \bibfield  [0]{\@secondoftwo}%
\providecommand \translation [1]{[#1]}%
\providecommand \BibitemOpen [0]{}%
\providecommand \bibitemStop [0]{}%
\providecommand \bibitemNoStop [0]{.\EOS\space}%
\providecommand \EOS [0]{\spacefactor3000\relax}%
\providecommand \BibitemShut  [1]{\csname bibitem#1\endcsname}%
\let\auto@bib@innerbib\@empty
%</preamble>
\bibitem [{\citenamefont {Hu}\ \emph {et~al.}(2006)\citenamefont {Hu} \emph
  {et~al.}}]{Hu:2006fy}%
  \BibitemOpen
  \bibfield  {author} {\bibinfo {author} {\bibfnamefont {B.}~\bibnamefont {Hu}}
  \emph {et~al.},\ }\href@noop {} {\bibfield  {journal} {\bibinfo  {journal}
  {Phys. Rev.}\ }\textbf {\bibinfo {volume} {C73}},\ \bibinfo {pages} {064004}
  (\bibinfo {year} {2006})}\BibitemShut {NoStop}%
\bibitem [{\citenamefont {MacLachlan}\ \emph {et~al.}(2006)\citenamefont
  {MacLachlan} \emph {et~al.}}]{MacLachlan:2006vw}%
  \BibitemOpen
  \bibfield  {author} {\bibinfo {author} {\bibfnamefont {G.}~\bibnamefont
  {MacLachlan}} \emph {et~al.},\ }\href@noop {} {\bibfield  {journal} {\bibinfo
   {journal} {Nucl. Phys.}\ }\textbf {\bibinfo {volume} {A764}},\ \bibinfo
  {pages} {261} (\bibinfo {year} {2006})}\BibitemShut {NoStop}%
\bibitem [{\citenamefont {Gayou}\ \emph {et~al.}(2001)\citenamefont {Gayou}
  \emph {et~al.}}]{Gayou:2001qt}%
  \BibitemOpen
  \bibfield  {author} {\bibinfo {author} {\bibfnamefont {O.}~\bibnamefont
  {Gayou}} \emph {et~al.},\ }\href@noop {} {\bibfield  {journal} {\bibinfo
  {journal} {Phys. Rev.}\ }\textbf {\bibinfo {volume} {C64}},\ \bibinfo {pages}
  {038202} (\bibinfo {year} {2001})}\BibitemShut {NoStop}%
\bibitem [{\citenamefont {Punjabi}\ \emph {et~al.}(2005)\citenamefont {Punjabi}
  \emph {et~al.}}]{Punjabi:2005wq}%
  \BibitemOpen
  \bibfield  {author} {\bibinfo {author} {\bibfnamefont {V.}~\bibnamefont
  {Punjabi}} \emph {et~al.},\ }\href@noop {} {\bibfield  {journal} {\bibinfo
  {journal} {Phys. Rev.}\ }\textbf {\bibinfo {volume} {C71}},\ \bibinfo {pages}
  {055202} (\bibinfo {year} {2005})}\BibitemShut {NoStop}%
\bibitem [{\citenamefont {Jones}\ \emph {et~al.}(2006)\citenamefont {Jones}
  \emph {et~al.}}]{Jones:2006kf}%
  \BibitemOpen
  \bibfield  {author} {\bibinfo {author} {\bibfnamefont {M.~K.}\ \bibnamefont
  {Jones}} \emph {et~al.},\ }\href@noop {} {\bibfield  {journal} {\bibinfo
  {journal} {Phys. Rev.}\ }\textbf {\bibinfo {volume} {C74}},\ \bibinfo {pages}
  {035201} (\bibinfo {year} {2006})}\BibitemShut {NoStop}%
\bibitem [{\citenamefont {Puckett}\ \emph {et~al.}(2010)\citenamefont {Puckett}
  \emph {et~al.}}]{Puckett:2010ac}%
  \BibitemOpen
  \bibfield  {author} {\bibinfo {author} {\bibfnamefont {A.~J.~R.}\
  \bibnamefont {Puckett}} \emph {et~al.},\ }\href@noop {} {\bibfield  {journal}
  {\bibinfo  {journal} {Phys. Rev. Lett.}\ }\textbf {\bibinfo {volume} {104}},\
  \bibinfo {pages} {242301} (\bibinfo {year} {2010})}\BibitemShut {NoStop}%
\bibitem [{\citenamefont {Paolone}\ \emph {et~al.}(2010)\citenamefont {Paolone}
  \emph {et~al.}}]{Paolone:2010qc}%
  \BibitemOpen
  \bibfield  {author} {\bibinfo {author} {\bibfnamefont {M.}~\bibnamefont
  {Paolone}} \emph {et~al.},\ }\href@noop {} {\bibfield  {journal} {\bibinfo
  {journal} {Phys. Rev. Lett.}\ }\textbf {\bibinfo {volume} {105}},\ \bibinfo
  {pages} {072001} (\bibinfo {year} {2010})}\BibitemShut {NoStop}%
\bibitem [{\citenamefont {Puckett}\ \emph {et~al.}(2012)\citenamefont {Puckett}
  \emph {et~al.}}]{Puckett:2011xg}%
  \BibitemOpen
  \bibfield  {author} {\bibinfo {author} {\bibfnamefont {A.~J.~R.}\
  \bibnamefont {Puckett}} \emph {et~al.},\ }\href@noop {} {\bibfield  {journal}
  {\bibinfo  {journal} {Phys. Rev.}\ }\textbf {\bibinfo {volume} {85}},\
  \bibinfo {pages} {045203} (\bibinfo {year} {2012})}\BibitemShut {NoStop}%
\bibitem [{\citenamefont {Litt}\ \emph {et~al.}(1970)\citenamefont {Litt} \emph
  {et~al.}}]{Litt:1969my}%
  \BibitemOpen
  \bibfield  {author} {\bibinfo {author} {\bibfnamefont {J.}~\bibnamefont
  {Litt}} \emph {et~al.},\ }\href@noop {} {\bibfield  {journal} {\bibinfo
  {journal} {Phys. Lett.}\ }\textbf {\bibinfo {volume} {B31}},\ \bibinfo
  {pages} {40} (\bibinfo {year} {1970})}\BibitemShut {NoStop}%
\bibitem [{\citenamefont {Bartel}\ \emph {et~al.}(1973)\citenamefont {Bartel}
  \emph {et~al.}}]{Bartel:1973rf}%
  \BibitemOpen
  \bibfield  {author} {\bibinfo {author} {\bibfnamefont {W.}~\bibnamefont
  {Bartel}} \emph {et~al.},\ }\href@noop {} {\bibfield  {journal} {\bibinfo
  {journal} {Nucl. Phys.}\ }\textbf {\bibinfo {volume} {B58}},\ \bibinfo
  {pages} {429} (\bibinfo {year} {1973})}\BibitemShut {NoStop}%
\bibitem [{\citenamefont {Andivahis}\ \emph {et~al.}(1994)\citenamefont
  {Andivahis} \emph {et~al.}}]{Andivahis:1994rq}%
  \BibitemOpen
  \bibfield  {author} {\bibinfo {author} {\bibfnamefont {L.}~\bibnamefont
  {Andivahis}} \emph {et~al.},\ }\href@noop {} {\bibfield  {journal} {\bibinfo
  {journal} {Phys. Rev.}\ }\textbf {\bibinfo {volume} {D50}},\ \bibinfo {pages}
  {5491} (\bibinfo {year} {1994})}\BibitemShut {NoStop}%
\bibitem [{\citenamefont {Walker}\ \emph {et~al.}(1994)\citenamefont {Walker}
  \emph {et~al.}}]{Walker:1993vj}%
  \BibitemOpen
  \bibfield  {author} {\bibinfo {author} {\bibfnamefont {R.~C.}\ \bibnamefont
  {Walker}} \emph {et~al.},\ }\href@noop {} {\bibfield  {journal} {\bibinfo
  {journal} {Phys. Rev.}\ }\textbf {\bibinfo {volume} {D49}},\ \bibinfo {pages}
  {5671} (\bibinfo {year} {1994})}\BibitemShut {NoStop}%
\bibitem [{\citenamefont {Christy}\ \emph {et~al.}(2004)\citenamefont {Christy}
  \emph {et~al.}}]{Christy:2004rc}%
  \BibitemOpen
  \bibfield  {author} {\bibinfo {author} {\bibfnamefont {M.~E.}\ \bibnamefont
  {Christy}} \emph {et~al.},\ }\href@noop {} {\bibfield  {journal} {\bibinfo
  {journal} {Phys. Rev.}\ }\textbf {\bibinfo {volume} {C70}},\ \bibinfo {pages}
  {015206} (\bibinfo {year} {2004})}\BibitemShut {NoStop}%
\bibitem [{\citenamefont {Qattan}\ \emph {et~al.}(2005)\citenamefont {Qattan}
  \emph {et~al.}}]{Qattan:2004ht}%
  \BibitemOpen
  \bibfield  {author} {\bibinfo {author} {\bibfnamefont {I.~A.}\ \bibnamefont
  {Qattan}} \emph {et~al.},\ }\href@noop {} {\bibfield  {journal} {\bibinfo
  {journal} {Phys. Rev. Lett.}\ }\textbf {\bibinfo {volume} {94}},\ \bibinfo
  {pages} {142301} (\bibinfo {year} {2005})}\BibitemShut {NoStop}%
\bibitem [{\citenamefont {Guichon}\ and\ \citenamefont
  {Vanderhaeghen}(2003)}]{Guichon:2003qm}%
  \BibitemOpen
  \bibfield  {author} {\bibinfo {author} {\bibfnamefont {P.~A.~M.}\
  \bibnamefont {Guichon}}\ and\ \bibinfo {author} {\bibfnamefont
  {M.}~\bibnamefont {Vanderhaeghen}},\ }\href@noop {} {\bibfield  {journal}
  {\bibinfo  {journal} {Phys. Rev. Lett.}\ }\textbf {\bibinfo {volume} {91}},\
  \bibinfo {pages} {142303} (\bibinfo {year} {2003})}\BibitemShut {NoStop}%
\bibitem [{\citenamefont {Blunden}\ \emph {et~al.}(2003)\citenamefont
  {Blunden}, \citenamefont {Melnitchouk},\ and\ \citenamefont
  {Tjon}}]{Blunden:2003sp}%
  \BibitemOpen
  \bibfield  {author} {\bibinfo {author} {\bibfnamefont {P.~G.}\ \bibnamefont
  {Blunden}}, \bibinfo {author} {\bibfnamefont {W.}~\bibnamefont
  {Melnitchouk}}, \ and\ \bibinfo {author} {\bibfnamefont {J.~A.}\ \bibnamefont
  {Tjon}},\ }\href@noop {} {\bibfield  {journal} {\bibinfo  {journal} {Phys.
  Rev. Lett.}\ }\textbf {\bibinfo {volume} {91}},\ \bibinfo {pages} {142304}
  (\bibinfo {year} {2003})}\BibitemShut {NoStop}%
\bibitem [{\citenamefont {Chen}\ \emph {et~al.}(2004)\citenamefont {Chen} \emph
  {et~al.}}]{Chen:2004tw}%
  \BibitemOpen
  \bibfield  {author} {\bibinfo {author} {\bibfnamefont {Y.~C.}\ \bibnamefont
  {Chen}} \emph {et~al.},\ }\href@noop {} {\bibfield  {journal} {\bibinfo
  {journal} {Phys. Rev. Lett.}\ }\textbf {\bibinfo {volume} {93}},\ \bibinfo
  {pages} {122301} (\bibinfo {year} {2004})}\BibitemShut {NoStop}%
\bibitem [{\citenamefont {Afanasev}\ \emph {et~al.}(2005)\citenamefont
  {Afanasev} \emph {et~al.}}]{Afanasev:2005mp}%
  \BibitemOpen
  \bibfield  {author} {\bibinfo {author} {\bibfnamefont {A.~V.}\ \bibnamefont
  {Afanasev}} \emph {et~al.},\ }\href@noop {} {\bibfield  {journal} {\bibinfo
  {journal} {Phys. Rev.}\ }\textbf {\bibinfo {volume} {D72}},\ \bibinfo {pages}
  {013008} (\bibinfo {year} {2005})}\BibitemShut {NoStop}%
\bibitem [{\citenamefont {Blunden}\ \emph {et~al.}(2005)\citenamefont
  {Blunden}, \citenamefont {Melnitchouk},\ and\ \citenamefont
  {Tjon}}]{Blunden:2005ew}%
  \BibitemOpen
  \bibfield  {author} {\bibinfo {author} {\bibfnamefont {P.~G.}\ \bibnamefont
  {Blunden}}, \bibinfo {author} {\bibfnamefont {W.}~\bibnamefont
  {Melnitchouk}}, \ and\ \bibinfo {author} {\bibfnamefont {J.~A.}\ \bibnamefont
  {Tjon}},\ }\href@noop {} {\bibfield  {journal} {\bibinfo  {journal} {Phys.
  Rev.}\ }\textbf {\bibinfo {volume} {C72}},\ \bibinfo {pages} {034612}
  (\bibinfo {year} {2005})}\BibitemShut {NoStop}%
\bibitem [{\citenamefont {Kondratyuk}\ \emph {et~al.}(2005)\citenamefont
  {Kondratyuk}, \citenamefont {Blunden}, \citenamefont {Melnitchouk},\ and\
  \citenamefont {Tjon}}]{Kondratyuk:2005kk}%
  \BibitemOpen
  \bibfield  {author} {\bibinfo {author} {\bibfnamefont {S.}~\bibnamefont
  {Kondratyuk}}, \bibinfo {author} {\bibfnamefont {P.~G.}\ \bibnamefont
  {Blunden}}, \bibinfo {author} {\bibfnamefont {W.}~\bibnamefont
  {Melnitchouk}}, \ and\ \bibinfo {author} {\bibfnamefont {J.~A.}\ \bibnamefont
  {Tjon}},\ }\href@noop {} {\bibfield  {journal} {\bibinfo  {journal} {Phys.
  Rev. Lett.}\ }\textbf {\bibinfo {volume} {95}},\ \bibinfo {pages} {172503}
  (\bibinfo {year} {2005})}\BibitemShut {NoStop}%
\bibitem [{\citenamefont {Mo}\ and\ \citenamefont {Tsai}(1969)}]{Mo:1968cg}%
  \BibitemOpen
  \bibfield  {author} {\bibinfo {author} {\bibfnamefont {L.~W.}\ \bibnamefont
  {Mo}}\ and\ \bibinfo {author} {\bibfnamefont {Y.-S.}\ \bibnamefont {Tsai}},\
  }\href {\doibase 10.1103/RevModPhys.41.205} {\bibfield  {journal} {\bibinfo
  {journal} {Rev. Mod. Phys.}\ }\textbf {\bibinfo {volume} {41}},\ \bibinfo
  {pages} {205} (\bibinfo {year} {1969})}\BibitemShut {NoStop}%
%%CITATION = RMPHA,41,205;%%
\bibitem [{\citenamefont {Maximon}\ and\ \citenamefont
  {Tjon}(2000)}]{Maximon:2000hm}%
  \BibitemOpen
  \bibfield  {author} {\bibinfo {author} {\bibfnamefont {L.~C.}\ \bibnamefont
  {Maximon}}\ and\ \bibinfo {author} {\bibfnamefont {J.~A.}\ \bibnamefont
  {Tjon}},\ }\href@noop {} {\bibfield  {journal} {\bibinfo  {journal} {Phys.
  Rev.}\ }\textbf {\bibinfo {volume} {C62}},\ \bibinfo {pages} {054320}
  (\bibinfo {year} {2000})}\BibitemShut {NoStop}%
\bibitem [{\citenamefont {Bystritskiy}\ \emph {et~al.}(2007)\citenamefont
  {Bystritskiy}, \citenamefont {Kuraev},\ and\ \citenamefont
  {Tomasi-Gustafsson}}]{Bystritskiy:2006ju}%
  \BibitemOpen
  \bibfield  {author} {\bibinfo {author} {\bibfnamefont {{\relax Yu}.~M.}\
  \bibnamefont {Bystritskiy}}, \bibinfo {author} {\bibfnamefont {E.~A.}\
  \bibnamefont {Kuraev}}, \ and\ \bibinfo {author} {\bibfnamefont
  {E.}~\bibnamefont {Tomasi-Gustafsson}},\ }\href@noop {} {\bibfield  {journal}
  {\bibinfo  {journal} {Phys. Rev.}\ }\textbf {\bibinfo {volume} {C75}},\
  \bibinfo {pages} {015207} (\bibinfo {year} {2007})}\BibitemShut {NoStop}%
\bibitem [{\citenamefont {Kuraev}\ \emph {et~al.}(2008)\citenamefont {Kuraev},
  \citenamefont {Bytev}, \citenamefont {Bakmaev},\ and\ \citenamefont
  {Tomasi-Gustafsson}}]{Kuraev:2007dn}%
  \BibitemOpen
  \bibfield  {author} {\bibinfo {author} {\bibfnamefont {E.~A.}\ \bibnamefont
  {Kuraev}}, \bibinfo {author} {\bibfnamefont {V.~V.}\ \bibnamefont {Bytev}},
  \bibinfo {author} {\bibfnamefont {S.}~\bibnamefont {Bakmaev}}, \ and\
  \bibinfo {author} {\bibfnamefont {E.}~\bibnamefont {Tomasi-Gustafsson}},\
  }\href {\doibase 10.1103/PhysRevC.78.015205} {\bibfield  {journal} {\bibinfo
  {journal} {Phys. Rev.}\ }\textbf {\bibinfo {volume} {C78}},\ \bibinfo {pages}
  {015205} (\bibinfo {year} {2008})},\ \Eprint {http://arxiv.org/abs/0710.3699}
  {arXiv:0710.3699 [hep-ph]} \BibitemShut {NoStop}%
%%CITATION = ARXIV:0710.3699;%%
\bibitem [{\citenamefont {Rachek}\ \emph {et~al.}(2015)\citenamefont {Rachek}
  \emph {et~al.}}]{Rachek:2014fam}%
  \BibitemOpen
  \bibfield  {author} {\bibinfo {author} {\bibfnamefont {I.~A.}\ \bibnamefont
  {Rachek}} \emph {et~al.},\ }\href@noop {} {\bibfield  {journal} {\bibinfo
  {journal} {Phys. Rev. Lett.}\ }\textbf {\bibinfo {volume} {114}},\ \bibinfo
  {pages} {062005} (\bibinfo {year} {2015})}\BibitemShut {NoStop}%
\bibitem [{\citenamefont {Adikaram}\ \emph {et~al.}(2015)\citenamefont
  {Adikaram} \emph {et~al.}}]{Adikaram:2014ykv}%
  \BibitemOpen
  \bibfield  {author} {\bibinfo {author} {\bibfnamefont {D.}~\bibnamefont
  {Adikaram}} \emph {et~al.},\ }\href@noop {} {\bibfield  {journal} {\bibinfo
  {journal} {Phys. Rev. Lett.}\ }\textbf {\bibinfo {volume} {114}},\ \bibinfo
  {pages} {062003} (\bibinfo {year} {2015})}\BibitemShut {NoStop}%
\bibitem [{\citenamefont {Chen}\ \emph {et~al.}(2007)\citenamefont {Chen},
  \citenamefont {Kao},\ and\ \citenamefont {Yang}}]{Chen:2007ac}%
  \BibitemOpen
  \bibfield  {author} {\bibinfo {author} {\bibfnamefont {Y.-C.}\ \bibnamefont
  {Chen}}, \bibinfo {author} {\bibfnamefont {C.-W.}\ \bibnamefont {Kao}}, \
  and\ \bibinfo {author} {\bibfnamefont {S.-N.}\ \bibnamefont {Yang}},\
  }\href@noop {} {\bibfield  {journal} {\bibinfo  {journal} {Phys. Rev.}\
  }\textbf {\bibinfo {volume} {B652}},\ \bibinfo {pages} {269} (\bibinfo {year}
  {2007})}\BibitemShut {NoStop}%
\bibitem [{\citenamefont {Guttmann}\ \emph {et~al.}(2011)\citenamefont
  {Guttmann}, \citenamefont {Kivel}, \citenamefont {Meziane},\ and\
  \citenamefont {Vanderhaeghen}}]{Guttmann:2010au}%
  \BibitemOpen
  \bibfield  {author} {\bibinfo {author} {\bibfnamefont {J.}~\bibnamefont
  {Guttmann}}, \bibinfo {author} {\bibfnamefont {N.}~\bibnamefont {Kivel}},
  \bibinfo {author} {\bibfnamefont {M.}~\bibnamefont {Meziane}}, \ and\
  \bibinfo {author} {\bibfnamefont {M.}~\bibnamefont {Vanderhaeghen}},\
  }\href@noop {} {\bibfield  {journal} {\bibinfo  {journal} {Eur. Phys. Jour.}\
  }\textbf {\bibinfo {volume} {A47}},\ \bibinfo {pages} {1} (\bibinfo {year}
  {2011})}\BibitemShut {NoStop}%
\bibitem [{\citenamefont {Bernauer}\ \emph
  {et~al.}(2014{\natexlab{a}})\citenamefont {Bernauer} \emph
  {et~al.}}]{Bernauer:2013tpr}%
  \BibitemOpen
  \bibfield  {author} {\bibinfo {author} {\bibfnamefont {J.~C.}\ \bibnamefont
  {Bernauer}} \emph {et~al.} (\bibinfo {collaboration} {A1}),\ }\href@noop {}
  {\bibfield  {journal} {\bibinfo  {journal} {Phys. Rev.}\ }\textbf {\bibinfo
  {volume} {C90}},\ \bibinfo {pages} {015206} (\bibinfo {year}
  {2014}{\natexlab{a}})}\BibitemShut {NoStop}%
\bibitem [{\citenamefont {Schmidt}(2016)}]{Schmidt:2016aa}%
  \BibitemOpen
  \bibfield  {author} {\bibinfo {author} {\bibfnamefont {A.}~\bibnamefont
  {Schmidt}},\ }\href@noop {} {Ph.D. thesis},\ \bibinfo  {school}
  {Massachusetts Institute of Technology}, \bibinfo {address} {Cambridge,
  Massachusetts} (\bibinfo {year} {2016})\BibitemShut {NoStop}%
\bibitem [{\citenamefont {Milner}\ \emph {et~al.}(2014)\citenamefont {Milner},
  \citenamefont {Hasell}, \citenamefont {Kohl}, \citenamefont {Schneekloth}
  \emph {et~al.}}]{Milner:2013daa}%
  \BibitemOpen
  \bibfield  {author} {\bibinfo {author} {\bibfnamefont {R.}~\bibnamefont
  {Milner}}, \bibinfo {author} {\bibfnamefont {D.~K.}\ \bibnamefont {Hasell}},
  \bibinfo {author} {\bibfnamefont {M.}~\bibnamefont {Kohl}}, \bibinfo {author}
  {\bibfnamefont {U.}~\bibnamefont {Schneekloth}},  \emph {et~al.},\
  }\href@noop {} {\bibfield  {journal} {\bibinfo  {journal} {Nucl. Instr.
  Meth.}\ }\textbf {\bibinfo {volume} {A741}},\ \bibinfo {pages} {1} (\bibinfo
  {year} {2014})}\BibitemShut {NoStop}%
\bibitem [{\citenamefont {Bernauer}\ \emph
  {et~al.}(2014{\natexlab{b}})\citenamefont {Bernauer} \emph
  {et~al.}}]{Bernauer:2014pva}%
  \BibitemOpen
  \bibfield  {author} {\bibinfo {author} {\bibfnamefont {J.~C.}\ \bibnamefont
  {Bernauer}} \emph {et~al.},\ }\href@noop {} {\bibfield  {journal} {\bibinfo
  {journal} {Nucl. Instr. Meth.}\ }\textbf {\bibinfo {volume} {A755}},\
  \bibinfo {pages} {20} (\bibinfo {year} {2014}{\natexlab{b}})}\BibitemShut
  {NoStop}%
\bibitem [{\citenamefont {Hasell}\ \emph {et~al.}(2009)\citenamefont {Hasell}
  \emph {et~al.}}]{Hasell:2009zza}%
  \BibitemOpen
  \bibfield  {author} {\bibinfo {author} {\bibfnamefont {D.~K.}\ \bibnamefont
  {Hasell}} \emph {et~al.},\ }\href@noop {} {\bibfield  {journal} {\bibinfo
  {journal} {Nucl. Instr. Meth.}\ }\textbf {\bibinfo {volume} {A603}},\
  \bibinfo {pages} {247} (\bibinfo {year} {2009})}\BibitemShut {NoStop}%
\bibitem [{\citenamefont {Perez~Benito}\ \emph {et~al.}(2016)\citenamefont
  {Perez~Benito} \emph {et~al.}}]{Benito:2016cmp}%
  \BibitemOpen
  \bibfield  {author} {\bibinfo {author} {\bibfnamefont {R.}~\bibnamefont
  {Perez~Benito}} \emph {et~al.},\ }\href@noop {} {\bibfield  {journal}
  {\bibinfo  {journal} {Nucl. Instr. Meth.}\ }\textbf {\bibinfo {volume}
  {A826}},\ \bibinfo {pages} {6} (\bibinfo {year} {2016})}\BibitemShut
  {NoStop}%
\bibitem [{\citenamefont {Ates}(2014)}]{Ates:2014aa}%
  \BibitemOpen
  \bibfield  {author} {\bibinfo {author} {\bibfnamefont {O.}~\bibnamefont
  {Ates}},\ }\href@noop {} {Ph.D. thesis},\ \bibinfo  {school} {Hampton
  University}, \bibinfo {address} {Hampton, Virginia} (\bibinfo {year}
  {2014})\BibitemShut {NoStop}%
\bibitem [{\citenamefont {Thiel}\ \emph {et~al.}(2012)\citenamefont {Thiel}
  \emph {et~al.}}]{Thiel:2012yj}%
  \BibitemOpen
  \bibfield  {author} {\bibinfo {author} {\bibfnamefont {A.}~\bibnamefont
  {Thiel}} \emph {et~al.},\ }\href {\doibase 10.1103/PhysRevLett.109.102001,
  10.1103/PhysRevLett.110.169102, 10.1103/PhysRevLett.110.169101} {\bibfield
  {journal} {\bibinfo  {journal} {Phys. Rev. Lett.}\ }\textbf {\bibinfo
  {volume} {109}},\ \bibinfo {pages} {102001} (\bibinfo {year} {2012})},\
  \Eprint {http://arxiv.org/abs/1207.2686} {arXiv:1207.2686 [nucl-ex]}
  \BibitemShut {NoStop}%
%%CITATION = ARXIV:1207.2686;%%
\bibitem [{\citenamefont {Bernauer}\ \emph {et~al.}(2016)\citenamefont
  {Bernauer} \emph {et~al.}}]{Bernauer:2016cc}%
  \BibitemOpen
  \bibfield  {author} {\bibinfo {author} {\bibfnamefont {J.~C.}\ \bibnamefont
  {Bernauer}} \emph {et~al.},\ }\href@noop {} {\bibfield  {journal} {\bibinfo
  {journal} {Nucl. Instr. Meth.}\ }\textbf {\bibinfo {volume} {A823}},\
  \bibinfo {pages} {9} (\bibinfo {year} {2016})}\BibitemShut {NoStop}%
\bibitem [{\citenamefont {Russell}(2016)}]{Russell:2016aa}%
  \BibitemOpen
  \bibfield  {author} {\bibinfo {author} {\bibfnamefont {R.~L.}\ \bibnamefont
  {Russell}},\ }\href@noop {} {Ph.D. thesis},\ \bibinfo  {school}
  {Massachusetts Institute of Technology}, \bibinfo {address} {Cambridge,
  Massachusetts} (\bibinfo {year} {2016})\BibitemShut {NoStop}%
\bibitem [{\citenamefont {Gramolin}\ \emph {et~al.}(2014)\citenamefont
  {Gramolin} \emph {et~al.}}]{Gramolin:2014pva}%
  \BibitemOpen
  \bibfield  {author} {\bibinfo {author} {\bibfnamefont {A.~V.}\ \bibnamefont
  {Gramolin}} \emph {et~al.},\ }\href@noop {} {\bibfield  {journal} {\bibinfo
  {journal} {J. Phys.}\ }\textbf {\bibinfo {volume} {G41}},\ \bibinfo {pages}
  {115001} (\bibinfo {year} {2014})}\BibitemShut {NoStop}%
\bibitem [{\citenamefont {Henderson}(2016)}]{Henderson:2016aa}%
  \BibitemOpen
  \bibfield  {author} {\bibinfo {author} {\bibfnamefont {B.~S.}\ \bibnamefont
  {Henderson}},\ }\href@noop {} {Ph.D. thesis},\ \bibinfo  {school}
  {Massachusetts Institute of Technology}, \bibinfo {address} {Cambridge,
  Massachusetts} (\bibinfo {year} {2016})\BibitemShut {NoStop}%
\bibitem [{\citenamefont {Bernauer}(2016)}]{Bernauer:analysis}%
  \BibitemOpen
  \bibfield  {author} {\bibinfo {author} {\bibfnamefont {J.~C.}\ \bibnamefont
  {Bernauer}},\ }\href@noop {} {\enquote {\bibinfo {title} {Elastic event
  selection},}\ } (\bibinfo {year} {2016}),\ \bibinfo {note}
  {unpublished}\BibitemShut {NoStop}%
\bibitem [{\citenamefont {O'Connor}(2017)}]{O'Connor:2016aa}%
  \BibitemOpen
  \bibfield  {author} {\bibinfo {author} {\bibfnamefont {C.}~\bibnamefont
  {O'Connor}},\ }\href@noop {} {Ph.D. thesis},\ \bibinfo  {school}
  {Massachusetts Institute of Technology}, \bibinfo {address} {Cambridge,
  Massachusetts} (\bibinfo {year} {2017})\BibitemShut {NoStop}%
\bibitem [{\citenamefont {Ice}(2016)}]{Ice:2016aa}%
  \BibitemOpen
  \bibfield  {author} {\bibinfo {author} {\bibfnamefont {L.~D.}\ \bibnamefont
  {Ice}},\ }\href@noop {} {Ph.D. thesis},\ \bibinfo  {school} {Arizona State
  University}, \bibinfo {address} {Tempe, Arizona} (\bibinfo {year}
  {2016})\BibitemShut {NoStop}%
\bibitem [{\citenamefont {Schmidt}\ \emph {et~al.}(2016)\citenamefont {Schmidt}
  \emph {et~al.}}]{Schmidt:MIE}%
  \BibitemOpen
  \bibfield  {author} {\bibinfo {author} {\bibfnamefont {A.}~\bibnamefont
  {Schmidt}} \emph {et~al.},\ }\href@noop {} {\bibfield  {journal} {\bibinfo
  {journal} {Nucl. Instr. Meth.}\ } (\bibinfo {year} {2016})},\ \bibinfo {note}
  {to be published}\BibitemShut {NoStop}%
\bibitem [{\citenamefont {Blunden}(2016)}]{Blunden:2016aa}%
  \BibitemOpen
  \bibfield  {author} {\bibinfo {author} {\bibfnamefont {P.}~\bibnamefont
  {Blunden}},\ }\href@noop {} {}\bibinfo {howpublished} {Private
  communication.} (\bibinfo {year} {2016})\BibitemShut {NoStop}%
\bibitem [{\citenamefont {Tomalak}\ and\ \citenamefont
  {Vanderhaeghen}(2015)}]{Tomalak:2014sva}%
  \BibitemOpen
  \bibfield  {author} {\bibinfo {author} {\bibfnamefont {O.}~\bibnamefont
  {Tomalak}}\ and\ \bibinfo {author} {\bibfnamefont {M.}~\bibnamefont
  {Vanderhaeghen}},\ }\href {\doibase 10.1140/epja/i2015-15024-1} {\bibfield
  {journal} {\bibinfo  {journal} {Eur. Phys. J.}\ }\textbf {\bibinfo {volume}
  {A51}},\ \bibinfo {pages} {24} (\bibinfo {year} {2015})},\ \Eprint
  {http://arxiv.org/abs/1408.5330} {arXiv:1408.5330 [hep-ph]} \BibitemShut
  {NoStop}%
%%CITATION = ARXIV:1408.5330;%%
\end{thebibliography}%

\end{document}